\documentclass[preprint,preprintnumbers,amsmath,amssymb]{revtex4}

\usepackage{graphicx}

\begin{document}

\title{Global Relationships in Fluctuation and Response in Adaptive Evolution}

\author{Chikara Furusawa$^{1*}$ and Kunihiko Kaneko$^{2*}$\\~\\}

\affiliation{$^{1}$Quantitative Biology Center, RIKEN, 6-2-3 Furuedai, Suita, Osaka 565-0874, Japan\\
$^{2}$Research Center for Complex Systems Biology,
Univ. of Tokyo, Komaba, Meguro-ku, Tokyo 153-8902, Japan
~\\~\\
$^{*}$correspondence:\\
chikara.furusawa@riken.jp and kaneko@complex.c.u-tokyo.ac.jp
}

\maketitle

\noindent
{\bf Abstract}\\
Cells generally change their internal state to adapt to an environmental change, and accordingly evolve in response to the new conditions. This process involves phenotypic changes that occur over several different time scales, ranging from faster environmental adaptation without a corresponding change in the genomic sequence to slower evolutionary dynamics involving genetic mutations and subsequent selection. In this regard, a question arises as to whether there are any relationships between such phenotypic changes over the different time scales at which adaptive evolution occurs. In this study, we analyzed simulated adaptive evolution in a simple cell model, and found that proportionality between concentration changes in adaptation and evolution over all components, and the proportion coefficients were closely linked to the change in the growth rate of a cell. Furthermore, we demonstrated that the phenotypic variances in component concentrations due to (non-genetic) noise and genomic alternations are proportional across all components. These global relationships in cellular states were also supported by phenomenological theory and transcriptome analysis of laboratory evolution in {\it Escherichia coli}. These findings provide a basis for the development of a quantitative theory of plasticity and robustness, and to determine the general restriction of phenotypic changes imposed by evolution.

\newpage

\section{Introduction}

When an environmental condition is changed, biological systems change their state to adapt and evolve to the environmental change. Despite the recognized importance to characterize the capacity of adaptation and evolution, discussions on evolvability and plasticity have thus far remained at a qualitative, rather than quantitative, level. On the other hand, the cellular internal state can now be quantitatively determined by measuring the abundances of a variety of components, including proteins and metabolites. Recent advances in high-throughput experimental analysis enable quantification of changes within such a high-dimensional state space \cite{omics_analysis}.

After an environmental change, cells may first respond by changing the abundances of cellular components without changing the genome sequence. The typical time scale of such environmental adaptation is generally shorter than several generations. On the other hand, over the long-term, i.e., over many generations, the internal state is gradually changed by evolutionary dynamics, in which the genome sequence is altered by mutations and individuals with higher fitness are generally selected. Indeed, experimental data of both changes in phenotype, reflecting changes in gene expression profiles, and changes in the genomic sequence throughout the course of evolution are now available; for example, much data are available from results obtained from the experimental evolution of {\sl E. coli} \cite{Lenski,Kishimoto,Suzuki}.

Therefore, an important question arises: is there a general relationship between short-term phenotypic changes in adaptation and long-term phenotypic changes in evolution?
Of course, the phenotypic changes that occur over different time scales are generally caused by different mechanisms, and thus the existence of any relationship between them would be non-trivial. However, it should be noted that the essence of cellular dynamics is reproduction, in which the abundance of each cellular component is roughly doubled, and this constraint imposed by cellular reproduction imposes a restriction on the time development in high-dimensional cellular state space. That is, it is possible that a (glivingh) cellular state would be restricted to a sub-space of the high-dimensional state space, described by a relatively smaller number of variables. Such restriction to low-dimensional dynamics can provide a non-trivial link between the phenotypic changes occurring in adaptation and the long-term changes occurring over the course of evolution \cite{Mu}. In fact, some studies have suggested that there is a common trend over the thousands of gene expression changes in adaptation and evolution, in which genes whose expressions exhibit a larger response to environmental change tend to also show a larger response in their expression at the evolutionary scale \cite{Ying,Marx}. Furthermore, global cellular behavior is represented by few macroscopic variables such as the growth rate and fitness of the system, which govern the entire (high-dimensional) dynamics in a cell. Therefore, it is important to uncover the possible relationships in the high-dimensional cellular dynamics between the expression of the thousands of proteins and metabolites and a macroscopic variable such as the growth rate, throughout the process of adaptation and evolution.

In addition to the phenotypic changes that occur after environmental changes during adaptation and evolution, the cellular state also generally exhibits fluctuations even under a constant environment and without genomic alternations, which originate from the stochastic nature of intra-cellular chemical reactions \cite{Elowitz,Log-normal}. The possible relationship between such non-genetic fluctuations and adaptive responses has also garnered much attention recently. The proportionality between such fluctuations and the evolutionary rate of fitness and phenotype has been demonstrated in bacterial experimental evolution and in simulations of toy cell models, which are supported by phenomenological theory \cite{Sato, KKFurusawa2005, KKPlosOne, ESB}, analogous to the proportionality between the fluctuation and response in statistical physics that has been well established since Einstein \cite{Einstein,Kubo}.
In the present context, evolution is considered the response to genetic change, and thus the proportionality between fluctuation and evolutionary rate means that components that are more variable by noise are also more variable by genetic change.

Considering the suggested proportionality between the response of cellular states to the environmental change and to genetic change, and also between the response and fluctuations, one may expect the existence of proportionality among two-by-two quantities, namely, fluctuations and responses induced by environmental (non-genetic) perturbations (noise) and by genetic changes (mutation). This grand relationship, if confirmed, is of critical importance to evolutionary biology, as it would provide a theoretical basis for the quantitative study on the plasticity and robustness underlying adaptive evolution, and could also set a general restriction as to the extent of phenotypic changes possible through (future) evolution. 
Indeed, the quantities constituting the relationship can now be measured over high-dimensional cellular dynamics, reflected as changes in the expression of thousands of genes.
However, thus far, experimental confirmation of this grand relationship remains premature, and is in need of further scrutiny.
Accordingly, at this stage, it is important to examine such a relationship by adopting an integrative approach combining {\sl in silico} evolution of a cell model consisting of thousands of chemical species, laboratory evolution of bacteria under environmental stress, and phenomenological theory for time development in a high-dimensional state space. In the present study, we aimed to uncover such statistical laws underlying the fluctuation and response of high-dimensional state variables occurring through adaptive evolution, and to connect them with changes in growth rate or fitness.

\section{Results}

\noindent
{\bf Evolutionary simulations of a simple cell model}

We employed our previously established mutually catalytic reaction network model, as this model is capable of capturing the basic characteristic of cells such as the power-law abundances, log-normal fluctuations, fluctuation-response relationship of fitness, adaptation with fold-change detection, and so forth, in spite of its simplicity \cite{Zipf, Log-normal, SOC-Zipf, KKFurusawa2005}.
In the model, the cellular state is represented by a set of molecule numbers $(N_1,N_2,\cdots,N_K)$, where $N_i$ is the number of molecules of the chemical species $i$, which ranges from $i=1$ to $K$.
For the internal chemical reaction dynamics, we chose a catalytic network among these $K$ chemical species, where each reaction from some chemical $i$ to some other chemical $j$ is catalyzed by a third chemical $\ell$.
Some resources (nutrients) are supplied from the environment by transportation through the cell membrane with the aid of some other chemicals that are termed 'transporters'.
The environmental condition is given by the concentrations of nutrient chemicals. Through catalytic reactions, these nutrients are transformed into cell-component chemicals, and a cell divides when the amount of component chemicals reaches a certain threshold.
Here, to achieve a higher growth rate, the synthesis of the cell components, transporters, and chemicals that catalyze the synthesis of those components need to be harmonized with the nutrient uptake.
We allowed the above toy-cell model consisting of catalytic reaction networks to evolve by rewiring the network paths with a given mutation rate and selecting the pathways with a certain fraction of cells that showed a higher growth rate (See Methods for details).
For a given environmental condition, evolution progresses so that the cell growth rate, i.e., the inverse of the average division time, is increased (Fig. 1).
To study the response to environmental change, we then switched the nutrient condition after evolution under a fixed condition for 3000 generations (denoted by the arrow in Fig. 1). The growth rate initially decreased following this environmental change, and then recovered through genetic evolution over generations. Next, we explored the phenotypic state changes in response to the environmental and evolutionary changes in order to evaluate the relationship between non-genetic and genetic responses upon environmental change.

As phenotypic state variables for the cell, we computed the abundances of each chemical $N_i$ at the division event. Here, it is convenient to choose $X_i=\log N_i$ as a phenotypic variable, since the abundance generally increases exponentially over time through cellular growth, and perturbation in a network is also generally amplified exponentially. Indeed, this choice of logarithmic abundances is also relevant to the theoretical argument presented below, as well as to transcriptome analysis of gene expression.
Note also that the abundances $N_i$ are distributed by cells, even for those sharing the same reaction network, due to stochasticity in reaction dynamics; thus, the average abundance over all cells is required to study the mean response of cells, denoted by $\langle \cdots \rangle$.

After the change in nutrient condition, the abundances of all the components change. Let us denote the average change of these abundances by:
$\delta X^{Env}_i \equiv \langle X_i(1)\rangle-\langle X_i(0) \rangle=\log \frac{\langle N_i(1) \rangle}{\langle N_i(0) \rangle}$, where generation 1 refers to the time point immediately following the environmental change, and generation 0 denotes the generation right before this nutrient change.
Similarly, we define the response by genetic evolution after $m$ generations by $\delta X^{Gen}_i(m) = \langle X_i(m) \rangle- \langle X_i(0) \rangle$. Fig. 2 shows the plot of
$\delta X^{Env}_i$ versus $\delta X^{Gen}_i(m)$ for $m=5$, 10, and 50. Interestingly, proportionality was found between the environmental and genetic responses over all components.

Let us now define this proportion coefficient $r(m)$ for $\frac{\delta X^{Gen}_i(m)}{\delta X^{Env}_i}$ across components $i$. This proportion coefficient $r(m)$ is initially close to 1, but with the increase in generations $m$, it decreases towards zero, in conjunction with the recovery of the growth rate. In other words, evolution shows a common tendency to abolish the changes in components introduced by the environmental change.
This common proportionality across all chemicals suggests that the proportion coefficient $r(m)$ is a ``global variable'' over a huge number of chemical species.
A reasonable candidate for such a global variable is the cell growth rate $\mu$. Hence, it is natural to compare the coefficient $r(m)$ with the growth rate. Toward this end, we again computed the change in the growth rate $\delta \mu^{Env}=\mu(1)-\mu(0)(<0)$ and $\delta \mu^{Gen}(m) =\mu(m)-\mu(0)$ at the $m$ th generation.
The ratio $\delta \mu^{Gen}(m)/\delta \mu^{Env}$ gives an index for the recovery in this growth rate from the decrease caused by the environmental change, with 0 and 1 representing full and null recovery, respectively. In Fig. 3, the proportion coefficient $r(m)$ is plotted against this growth rate recovery $\delta \mu^{Gen}(m)/\delta \mu^{Env}$. The proportionality between the two is clearly discernible.

Recalling the possible relationship between fluctuation and response, as is typical in statistical physics, we then evaluated whether there also exists a common relationship among the variances of all the components.
Here, as previously reported \cite{Log-normal}, the distribution of each $N_i$ follows an approximately log-normal distribution, as confirmed experimentally in the protein abundances in the present cells. Hence, it is again relevant to adopt $X_i=\log N_i$ as a phenotype variable, so that the distribution of $X_i$ follows a roughly Gaussian distribution \cite{Sato}. The phenotypic variance $V_{ip}(i)$ for each component $i$ is defined as the variance of $X_i$ in an isogenic population.
On the other hand, the variance due to genetic change $V_g(i)$ is defined as the variance of mean $X_i$ over a heterogenic distribution, where the mean is computed across clones of a given genotype (i.e., a network), while the heterogenic distribution is related to different genotypes (networks) that exist at a given generation.

The results of simulations are given in Fig. 4, which shows proportionality between $V_{ip}(i)$ and $V_g(i)$ across the components $i$ for the evolved population.
As the mutation rate is increased over evolution, $V_g(i)$ increases as the genotype distribution is broadened, whereas $V_{ip}(i)$ remains at the same level, so that the ratio $V_g(i)/V_{ip}(i)$ is increased while maintaining the proportionality.
This observed proportionality means that the components that are more variable owing to noise in the reaction dynamics are also more variable owing to mutation.

So far, we have confirmed the existence of common proportionality between non-genetic and genetic variances, as well as between the environmental and evolutionary responses.
As the proportionality between the fluctuation and response is a natural outcome in statistical physics, we compared the response and fluctuations in more detail.
However, direct comparison of the phenotypic variance $V_{ip}(i)$ with the environmental response did not show a clearly discernable proportionality. This is probably due to the discrepancy in the definitions of the two quantities: the variance originates from very high-dimensional dynamics without any specific directional change, while in the environmental response, only one specific environmental change considering only a few nutrients is applied. To make a more direct comparison, we then sampled the environmental responses against a variety of external changes introduced by different nutrient conditions to define the average environmental response $R^{env}(i) =\langle (\delta X^{env})^2 \rangle$ with $\langle \cdots \rangle$ over $10^4$ environmental conditions (see Methods). As shown in Fig. S1, this average environmental response showed clear proportionality with $V_{ip}(i)$ and $V_g(i)$, respectively. Hence, the proportionality relationships among two-by-two quantities, i.e., genetic and non-genetic responses and fluctuations, hold over all components (see Fig. 5).

Thus, the degree of plasticity required to achieve an adaptive response to a new environment is characterized by fluctuations $V_{ip}(i)$, i.e., those that do not consider environmental or genetic changes.
On the other hand, when cells are exposed to a novel environment, the potential of adaptation is expected to increase.
Therefore, when placed in a novel condition, it is expected that the phenotypic fluctuations would increase to allow the cells to adapt to the new environment. In Fig. S2, the variances $(V_{ip}(i), V_g(i))$ are plotted before and after the environmental change (arrow in Fig. 1). In this case, all of the variances increase while roughly maintaining their proportionality.
After this increase, the variances decrease over generations under a fixed environmental condition, while the proportionality between $V_{ip}(i)$ and $V_g(i)$ is maintained.

~\\
\noindent
{\bf Theoretical Argument}

By using a simple cell model, we have confirmed the common proportionality over thousands of components for genetic and non-genetic variances, and environmental and genetic responses.
The results suggest the existence of a global variable that governs adaptive evolution. Here, the growth rate $\mu$ of a cell is a candidate for such a variable, since, for a cell to maintain its composition, every component has to be synthesized in conjunction with the growth rate. Indeed, in \cite{Mu} we considered the dynamics of gene expression
\begin{equation}
dx_i/dt=f_i(\{x_j\}) -\mu x_i,
\end{equation}
where $\mu x_i$ gives the dilution of the concentration by the increase in cell volume $V$, and $x_i$ is the concentration of the component $i$, $x_i=N_i/V$. By using $ X_i =\log x_i$ and $F_i(\{ X_j\})x_i=f_i(\{x_j\})$, the original stationary state is given by
\begin{equation}
F_i(\{X^*_j \}) =\mu.
\end{equation}
Now, with the change in environmental condition $E$ and genetic change $G$, the expression $X_i$ is shifted to $X^*_i + \delta X_i$, and $\mu$ is shifted to $\mu+\delta\mu$. Assuming that the change in logarithmic concentration is not so large, and taking only the linear part of the changes and using
the Jacobi matrix $J_{ij}=(\frac{\partial F_i}{\partial X_j})_{X_m=X_m^*}$, we get
\begin{equation}
\sum_j J_{ij} \delta X_j (E,G) + \gamma _i^E \delta E +\gamma_i^G \delta G=\delta \mu (E,G),
\end{equation}
where
$\gamma_i^{E} \equiv \frac{\partial F_i}{\partial E}$ and $\gamma_i^{G} \equiv \frac{\partial F_i}{\partial G}$, respectively. 

Here, $G$ is a coordinate introduced to represent the genetic change.
It is not evident that the genetic change is represented by only a single variable. However, considering that under this scenario evolution progresses under a stressed environmental condition, one could project high-dimensional genetic change in the direction required to increase fitness (growth rate) under the condition, indicating that a single variable $G$ can be introduced; indeed, several studies conducted to date support this assumption \cite{Sato,KKFurusawa2005,ESB}.
Accordingly, the variable $G$ has the same dimensions as $E$, and can be scaled so that $G$ and $E$ induce the same degree of change in expression.
The genetic evolution following the initial stress $\delta E$ is expected to diminish the environmental stresses, so that evolution occurs in the direction $\delta G<0$, if the environmental change $\delta E$ is positive.
Considering that evolution occurs through the projected direction in $\delta E$, it is natural to assume $\gamma_i^E = \gamma_i^G$ (although this might be a crude approximation).
Under the linear conditions of interest, the change in $\mu$ is proportional to $\delta E$ or $\delta G$, with $\delta \mu (\delta E, \delta G)=\alpha (\delta E +\delta G)$. Note that, again, the direction of $\delta G$ is opposite to $\delta E$.
Thus, we obtain,
\begin{equation}
\delta X_j (\delta E,\delta G)= \sum_i L_{ji}(\delta \mu (\delta E,\delta G) - \gamma_i (\delta E+\delta G))
= \delta \mu (\delta E,\delta G) \sum_i L_{ji}(1- \gamma_i/\alpha).
\end{equation}
Then, over the course of evolution $\delta G=0$ to $\delta G(m)$, under a given environmental condition $E$,
\begin{equation}
\frac{\delta X_j^{Gen}(m)}{\delta X_j^{Env}}= \frac{\delta X_j(\delta E,\delta G(m))}{\delta X_j (\delta E,0)} =\frac{\delta \mu (\delta E,\delta G(m))}{\delta \mu (\delta E,0)}.
\end{equation}
In other words, all the expression changes are proportional, as confirmed in the present simulations. To check the validity of the theory, we compared the proportion coefficient in the expression change (LHS of eq.5) with the change in growth rate (RHS) numerically through the course of the evolution simulation. As shown in Fig. 3, the relationship of eq. 5 holds rather well.
Note that if there is deviation from $\gamma^E_i=\gamma^G_i$ over $i$, the proportionality over all genes will deviate. In other words, the deviation from $\delta X_j(\delta E,\delta G) \propto \delta X_j (\delta E,0) $ across genes $i$ reflects the deviation between $\gamma_j^E$ and $\gamma_j^G$.

The relationship in the variances $V_{ip}(i)$ and $V_g(i)$ is considered in a similar manner.
Consider that the fluctuation in $\delta E$ and $\delta G$ induce fluctuation in each expression $i$, induced by either noise or genetic variation. This fluctuation induces variation in the growth rate, according to $\delta \mu= \alpha \delta E$ or $\alpha \delta G$, so that
\begin{equation}
\langle (\delta X_j (\delta \Upsilon)^2 \rangle= \langle \delta \mu (\delta \Upsilon )^2\rangle (\sum_i L_{ji}(1 -\gamma_i/\alpha))^2,
\end{equation}
where $\delta \Upsilon$ is either $\delta E$ or $\delta G$, i.e., phenotypic change induced by variation in the environment (i.e., noise) or by genetic change (e.g., by mutation),
and $\langle \cdots \rangle$ is the average over the distribution induced by the phenotypic noise or genetic variation.
The variance $V_{ip}(j)$ and $V_g(j)$ are $\langle (\delta X_j(\delta E))^2\rangle$ and $\langle (\delta X_j(\delta G))^2\rangle $, respectively, so that

\begin{equation}
\frac{V_{ip}(j)}{V_g(j) } =\frac{\langle \delta \mu (\delta E)^2\rangle }{\langle \delta \mu (\delta G)^2\rangle }= \frac{V_{ip}(\mu)}{V_g(\mu)}.
\end{equation}
Thus, the ratio of the two variances takes on the same value independent of $j$, which is determined by the ratio of variances in growth rate fluctuations induced by noise to those induced by genetic variation.
This relationship was again confirmed in our simulated evolution model (see Fig. S3) (see also \cite{KKPlosOne} for an alternative derivation of the common proportionality between $V_{ip}(i)/V_g(i)$ assuming the common error of catastrophe in the phenotype distribution).

The above theoretical interpretations on both the proportionality in responses and in variances suggest the importance of changes in the growth rate.
Since the growth rate globally governs all of the concentrations through dilution, its dominance over each component is a reasonable assumption.
To further evaluate the relationship between environmental and evolutionary dynamics, however, we need to also assume that evolution progresses as to assimilate environmental change, as Waddington proposed \cite{Waddington}.
In our theory, this genetic assimilation is formulated by the introduction of the variable $G$ that has a similar effect with the environment (or compensates for the environmental stress), so that ${\partial F_i({x_j})}/{\partial E} \approx {\partial F_i({x_j})}/{\partial G}$.

~\\
\noindent
{\bf Experimental Verification}

The theoretical argument and the simulation results demonstrated the existence of a common proportion coefficient $r(m)$ for $\frac{\delta X^{Gen}_i(m)}{ \delta X^{Env}_i }$ across components $i$, and its proportionality to the growth rate recovery $\delta \mu^{Gen}(m)/\delta \mu^{Env}$.
To verify this relationship, we analyzed the time-series transcriptome data obtained in an experimental evolution study of {\sl E. coli} under conditions of ethanol stress \cite{Horinouchi1, Horinouchi2}. In this experiment, after cultivation of approximately 1,000 generations (2,500 hours) under 5\% ethanol stress, 6 independent ethanol-tolerant strains were obtained, which exhibited an approximately 2-fold increase in specific growth rates in comparison to the ancestor. For all independent culture series, mRNA samples were extracted from approximately $10^8$ cells at 6 different time points, and the absolute expression levels were quantified by using microarray analysis. All mRNA samples were obtained from the cells in exponential growth phase, which means that the changes in cellular state over the time scale of several generations were negligible, and each expression level represented cells in a steady-growth state (see \cite{Horinouchi2} for details of materials and methods).

Using the time-series expression data of bacterial adaptive evolution, we analyzed the common proportionality in expression changes. The environmental response of the $i$-th gene $\delta X^{Env}_i$ is defined by the log-transformed ratio of the expression level of the $i$-th gene obtained 24 hours after exposure to the stress condition to that obtained under the no-stress condition.
Similarly, the evolutionary response at $n$ hours after the exposure to the stress $\delta X^{Gen}_i (n) $ is defined by the log-transformed ratio of the expression level at $n$ hours to that of the non-stress condition.
We found a common trend between the environmental and genetic responses over all genes, as shown in Fig. 6(a).
Furthermore, we also found that the proportion coefficient $r(n)$ for ${\delta X^{Gen}_i(n)}/{ \delta X^{Env}_i }$ is roughly proportional to the growth recovery ratio $\delta \mu^{Gen}(n)/\delta \mu^{Env}$, as shown in Fig. 6(b), where $\delta\mu^{Gen}(n)$ and $\delta \mu^{Env}$ are the growth rate differences of $n$ hours and 24 hours after the exposure to stress, respectively.
The results demonstrated that the evolutionary dynamics with growth recovery were accompanied by gene expression changes to eliminate the phenotypic changes introduced by the new environment, and agreed well with the simulation results of the simple cell model shown in Fig. 2 as well as the theoretical argument presented above.

Furthermore, there is some indirect experimental support for the proposed relationships between the variances.
Stearns and colleagues measured the isogenic variance $V_{ip}$ of five life-history traits (such as body weight, lifespan, etc.) in {\sl Drosophila melanogaster}, as well as the genetic variance $V_g$ between different genetic lines observed during laboratory evolution to increase the traits, and observed proportionality between the two \cite{Stearns}.
The correlation between the isogenic variances in trait expression and variance due to mutation (but without selection) was measured across a few thousand genes in {\sl Saccharomyces cerevisiae}. Correlation between the two variances was observed \cite{Lehner}, whereas proportionality was not so clear. This is possibly because evolution without selection was applied in the experiment, and therefore only the variance resulting from random mutation was measured.

\section{Discussion}
We have shown proportionality in the change in the concentrations of most intra-cellular components as a result of adaptive evolution, which was confirmed in simulated evolution of catalytic reaction network models of cells, laboratory experiments of bacterial evolution, and phenomenological theory.
As the theoretical argument, albeit phenomenological, is rather general, we expect that the observed relationships obtained from the simulation and laboratory experiments represent a general phenomenon, independent of the specific models or organisms considered.
This proportionality across thousands of components implies that there is a strong constraint in phenotypic evolution.
In particular, the expression of different components cannot evolve independently, but rather change together, for the most part, along a one-dimensional path provided by eq.(5).
Phenotypic change in adaptive evolution under a fixed environmental condition is highly constrained; thus, we here quantitatively formulate the general restriction or feasibility of the direction of phenotype changes in future evolution.

Our result also implies that the changes in the concentrations of most components that are induced by the environmental change become relaxed through the evolutionary process.
This suggests a phenomenon of strong homeostasis, that is, a tendency to restore to the original, adapted, intra-cellular states, via genetic change.
In some sense, this homoeostasis is similar to the Le Chatelier principle in thermodynamics, in that changes introduced by external perturbations are relaxed by subsequent temporal evolution.

There could be a huge variety of genetic changes that yield the phenotypic changes required for adaptation.
Indeed, in our simulations, there were a variety of network structures that could achieve phenotypic adaptation.
When the simulation was run again with a different seed of random numbers for mutations, the resulting network (i.e., genotypes) was different in each run, but the change in concentrations (phenotypes) followed the proportionality given by eq.(5), independently of the specific genetic changes occurring during evolution.
Furthermore, in bacterial evolution experiments, the results from different strains tended to follow the same proportionality law described by eq.(5).
It is interesting to note that such correlated change in expression levels by genetic changes is also suggested in several experiments \cite{Ying,Marx}.
It will be important to further confirm the relationship eq.(5) in more laboratory evolution experiments, and to also unveil the underlying genotype-phenotype map that achieves the common, restricted change in expression levels observed in the experimental data.

We have also found proportionality in the fluctuations in expression levels across components.
As expected from Fisher's fundamental theorem of natural selection \cite{Fisher}, the higher the genetic variance, the higher the evolutionary rate.
Hence, the proportionality between $V_{ip}(i) \propto V_g(i)$ suggests that a higher isogenic variance of a given expression level due to noise would be accompanied by a higher rate in the change in the expression level due to evolution.
Hence, our results suggest that the direction of evolutionary change in phenotypic space is likely to be predetermined by the isogenic variance of expression level due to noise.

According to our theoretical framework, the responses and fluctuations in expression levels are represented by the macroscopic growth rate and its fluctuation.
Therefore, the relationship between the response and fluctuations, analogous to thermodynamics, is represented by the landscape of the growth rate as a function of phenotype (expression level) and the environment, in contrast to the established fitness landscape represented in genetic space proposed by Sewall Wright \cite{Wright}.
We hope that the present study will provide a basis for the development of a future macroscopic theory for phenotypic evolution.

\section{Methods: model simulations}

The cellular state can be represented by a set of numbers $(N_1,N_2,\cdots,N_K)$, where
$N_i$ is the number of molecules of the chemical species $i$ with $i$ ranging from $i=1$ to $K$.
For the internal chemical reaction dynamics, we chose a catalytic network among these $k$ chemical species, where each reaction from some chemical $i$ to some other chemical $j$ is assumed to be catalyzed by a third chemical $\ell$, i.e., $(i + \ell \rightarrow j + \ell)$.
A catalytic network is chosen randomly such that the probability that any two chemicals, $i$ and $j$, are connected is given by the connection rate $\rho$.
Some resources (nutrients) are supplied from the environment by transportation through the cell membrane with the aid of some other chemicals that are named `transporters'.
The concentrations of nutrient chemicals in the environment are kept constant, and they have no catalytic activity in order to prevent the occurrence of catalytic reactions in the environment.
Through the catalytic reactions, these nutrients are transformed into other chemicals, including the transporters.
Here, we assume that all of the K chemical species are necessary for cell division.
Thus, cell division is assumed to occur when the minimum number of species exceeds a threshold $M$, i.e., $\min_{i=1}^{K} N_i \geq M$ (in all analyses $M$ is set to unity).
Chosen randomly, the parent cell's molecules are evenly split among the two daughter cells.
In our numerical simulations, we randomly picked up a pair of molecules in a cell and transformed them according to the reaction network.
In the same way, transportation through the membrane was also computed by randomly choosing from molecules within the cell and from nutrients in the environment.
The parameters were set as $K=1000$ and $\rho=0.01$.

We studied the evolution of the replication dynamics by generating slightly modified networks and selecting those that grew faster. First, $n$ parent cells were generated, and the connecting paths of catalytic networks were chosen randomly with connection rate $\rho$.
From each of the $n$ parent cells, $L$ mutant cells were generated by randomly replacing $m \rho K^2$ reaction paths, where $\rho K^2$ is the total number of reactions and $m$ is the mutation rate per reaction per generation.
Then, reaction dynamics were simulated for each of the $nL$ cells to determine the rate of growth of each cell; that is, the inverse of the time required for division. Within the cell population, $n$ cells with faster growth rates were selected to be the parent cells of the next generation, from which $nL$ mutant cells were again generated in the same manner.
Throughout the simulation, the parameters were set as $n = 1000$ and $L = 5$.
In the simulations shown in Fig. 1, the mutation rate $m$ was set to $1 \times 10^{-3}$.

The environmental change is given by changing the nutrient concentration ratio in the environment.
In the evolutionary simulation shown in Fig. 1, there are two nutrient chemicals, each associated with one transporter chemical.
Before adding the new environmental condition (generation$\le$0), the concentrations of these two nutrients in the environment $(c_1, c_2)$ were set to $(0.5,0.5)$, while after the environmental change (generation$>$0), they were set to $(0.9, 0.1)$.
In the result shown in Fig. S1, to add a variety of environmental changes, we randomly selected a nutrient chemical and a transporter chemical for this nutrient among $K$ total chemical species.
Then, the concentrations of the new nutrient $c_{New}$ and the original nutrients were set to $(c_1, c_2, c_{New}) = (0.45, 0.45, 0.1)$.
We iterated the random addition of a nutrient $10^4$ times to obtain the average environmental response $R^{env}(i)$.

~\\
\noindent
{\bf Acknowledgements}\\

\noindent
{\bf General:} The authors would like to thank T. Yomo for stimulating discussions and constructive comments.\\

\noindent
{\bf Funding:} This work was supported in part by the platform for Dynamic Approaches to the Living Systems from the Ministry of Education, Culture, Sports, Science and Technology (MEXT), Japan. This work was also supported in part by Grant-in-Aid for Scientific Research on Innovative Areas [25128715 and 26119719 to C.F.] from MEXT, Japan.\\

\noindent
{\bf Author contribution:} C.F. and K.K. designed the study. C.F. performed the simulations and 
data analysis. K.K. performed the theoretical analysis. C.F. and K.K. wrote the manuscript. \\

\noindent
{\bf Competing financial interests:}
The authors declare no competing financial interests.\\
~\\

\noindent
{\bf Figure Captions}\\
~\\

\noindent
{\bf Figure 1}\\
Growth rate (fitness) increase over generations from generation $-3000$ to 0. At generation 0, the nutrient concentrations in the environment were changed. As a result, the growth rate drastically decreased at generation 0, and later recovered by the evolutionary dynamics (see Methods for details).

~\\
\noindent
{\bf Figure 2}\\
Response by environmental change versus response by evolution.
Relationship between the environmental response $\delta X^{Env}_i$ and genetic response $\delta X^{Gen}_i(m)$. (a), (b), and (c) show the plots for $m=5$, 10, and 50, respectively.
The black solid lines are $y=x$ for reference.

~\\
\noindent
{\bf Figure 3}\\
The relationship between growth recovery rate $\delta \mu^{Gen}(m)/\delta \mu^{Env}$ and the proportion coefficient $r(m)$. The proportion coefficient $r(m)$ was obtained by using the least-squares method for the relationship of $\delta X^{Env}_i$ and $\delta X^{Gen}_i(m)$ for $m=1 \sim 200$. The black solid line is $y=x$ for reference.

~\\
\noindent
{\bf Figure 4}\\
The relationship between $V_{ip}(i)$ and $V_g(i)$. The variances were computed by using the network and environment at generation 0 (before the environmental change) shown in Fig. S1 with various mutation rates.
$V_{ip}(i)$ and $V_g(i)$ were calculated based on the simulation results of randomly generated $10^5$ networks.
The solid line is $y=x$ for reference.

~\\
\noindent
{\bf Figure 5}\\
Proportionality relationships among genetic/non-genetic fluctuations and responses hold over all components. The arrows indicate the proportional relationships.

~\\
\noindent
{\bf Figure 6}\\
Response by environmental change versus response by evolution in {\sl E. coli} adaption to ethanol stress.
(a) Relationship between environmental response $\delta X^{Env}_i$ and genetic response $\delta X^{Gen}_i(n)$ for $n=2496$ as a representative example.
$\delta X^{Env}_i$ and $\delta X^{Gen}_i(n)$ were calculated by the log-transformed expression ratio between before and 24 hours after and $n$ hours after exposure to ethanol stress.
The blue line is obtained by least-squares fitting, while the black line is $y=x$ for reference.
(b) Relationship between growth recovery rate $\delta \mu^{Gen}(n)/\delta \mu^{Env}$ and the proportion coefficient $r(n)$.
The proportion coefficient $r(n)$ was obtained by using the least-squares method for the relationship of $\delta X^{Env}_i$ and $\delta X^{Gen}_i(n)$ for $n=384$, 744, 1224, 1824, and 2496 hours.
The growth recovery rate $\delta \mu^{Gen}(n)/\delta \mu^{Env}$ was calculated based on the experimental measurements (see \cite{Horinouchi2} for details).
Among the 6 independent culture lines in \cite{Horinouchi2}, the results of 5 culture lines without genome duplication are plotted.
The black line is $y=x$ for reference.

\newpage

\noindent
{\Large {\bf Figures}}

\begin{figure}[htbp]
\begin{center}
\includegraphics[width=12cm]{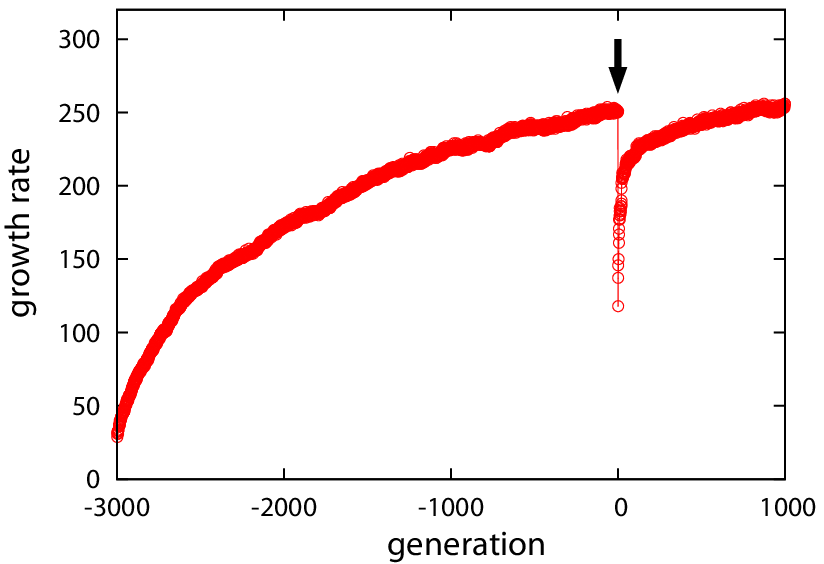}
\end{center}
\caption{
}
\end{figure}

\begin{figure}[htbp]
\begin{center}
\includegraphics[width=7cm]{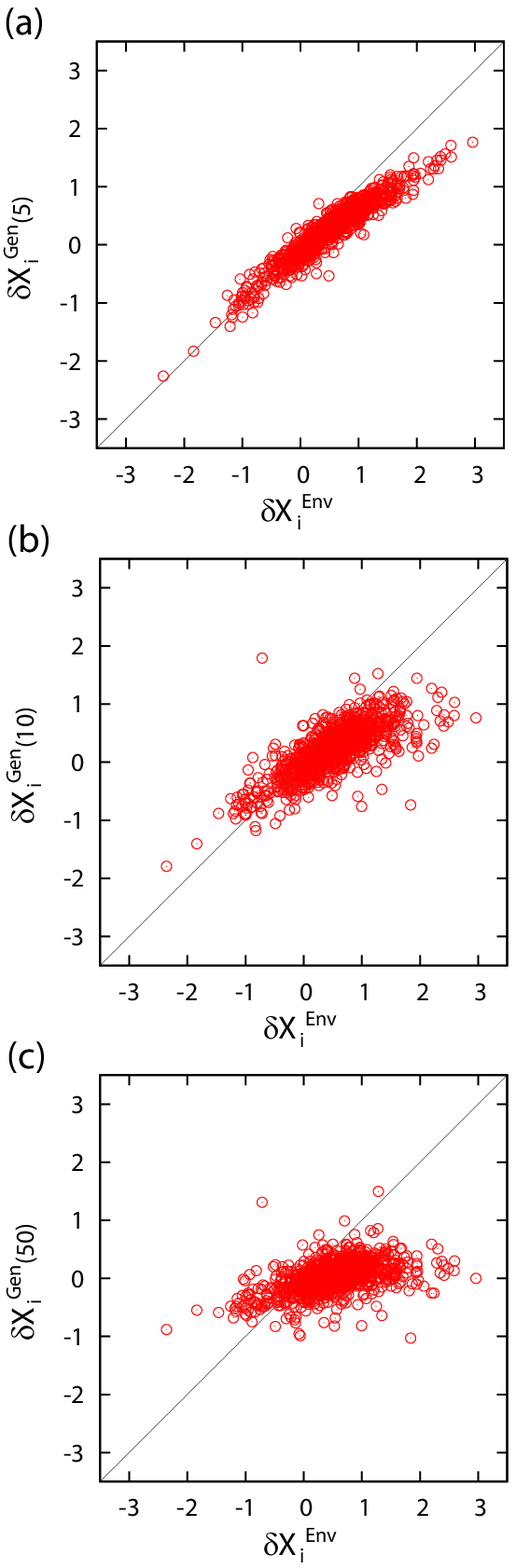}
\end{center}
\caption{
}
\end{figure}

\begin{figure}[htbp]
\begin{center}
\includegraphics[width=8cm]{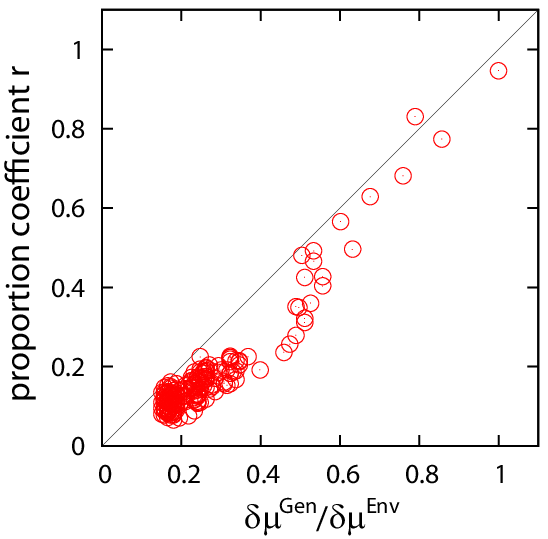}
\end{center}
\caption{
}
\end{figure}

\begin{figure}[htbp]
\begin{center}
\includegraphics[width=12cm]{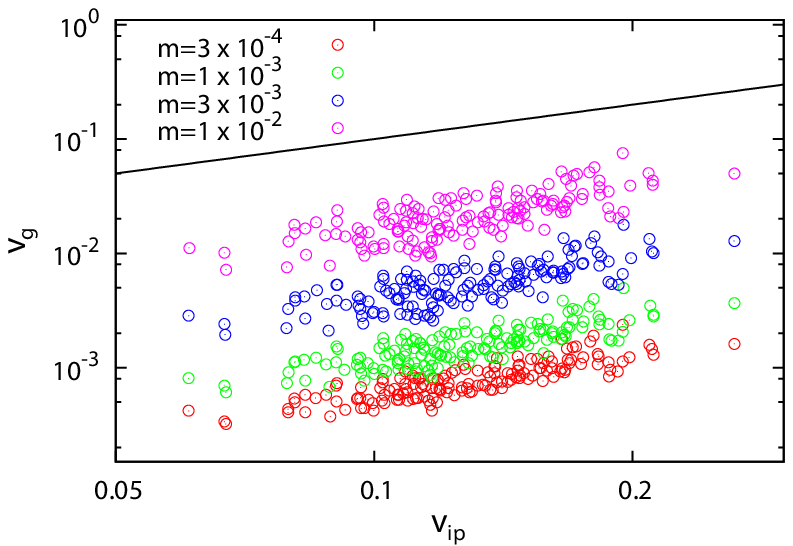}
\end{center}
\caption{
}
\end{figure}

\begin{figure}[htbp]
\begin{center}
\includegraphics[width=12cm]{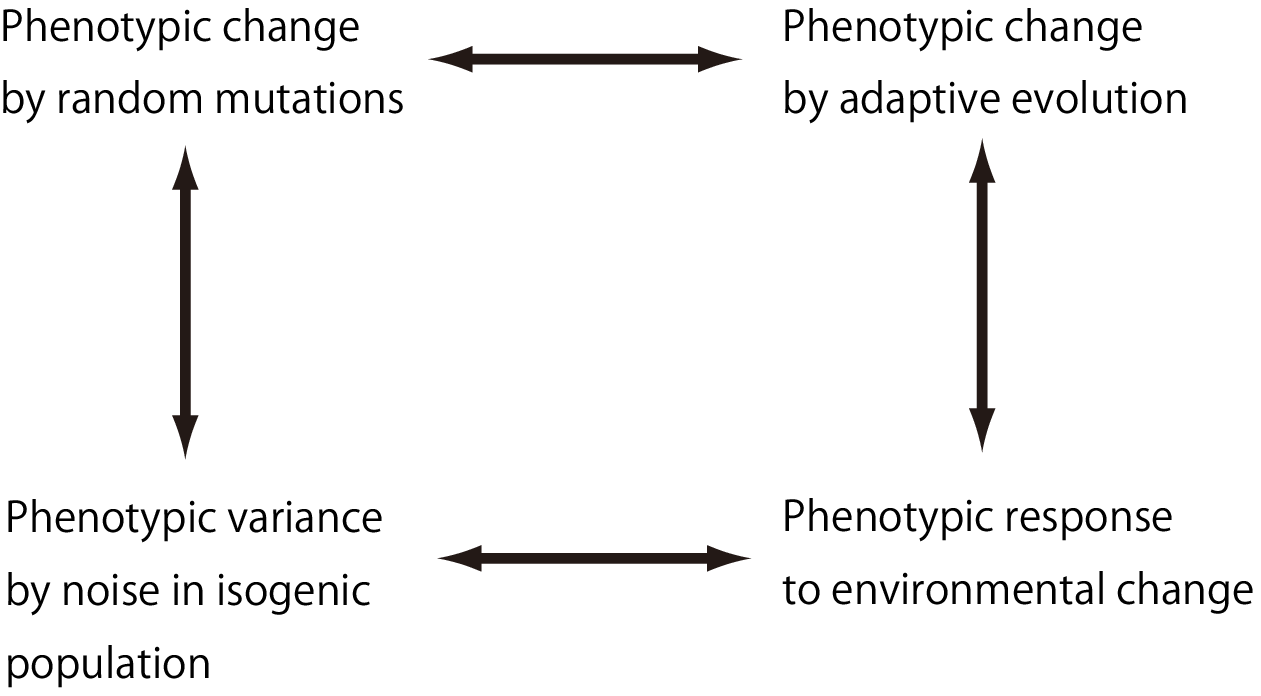}
\end{center}
\caption{
}
\end{figure}

\begin{figure}[t]
\begin{center}
\includegraphics[width=8cm]{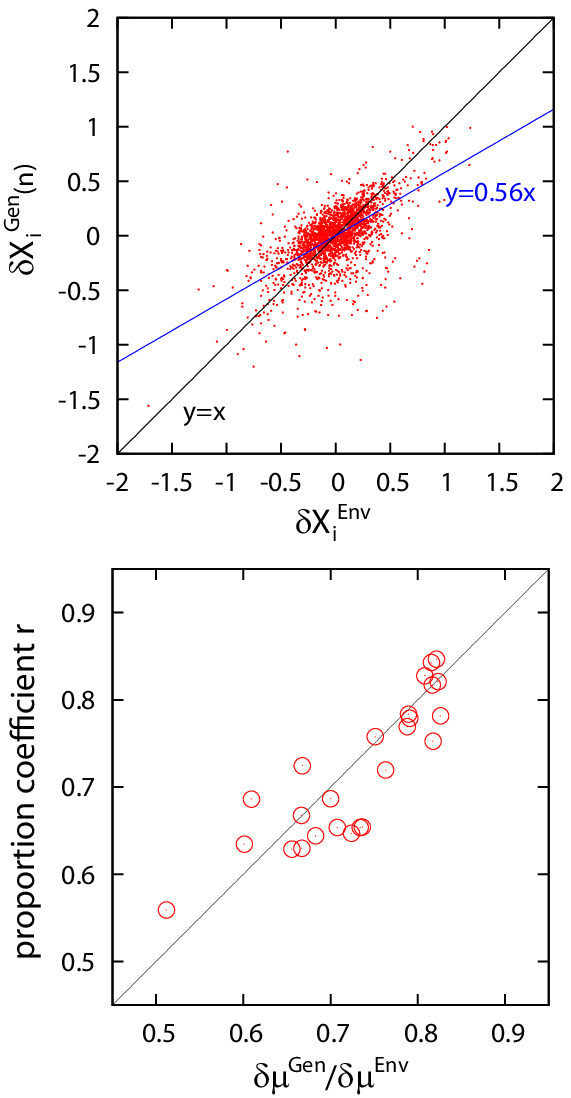}
\end{center}
\caption{
}
\end{figure}

\begin{figure}[htbp]
\begin{center}
\includegraphics[width=8cm]{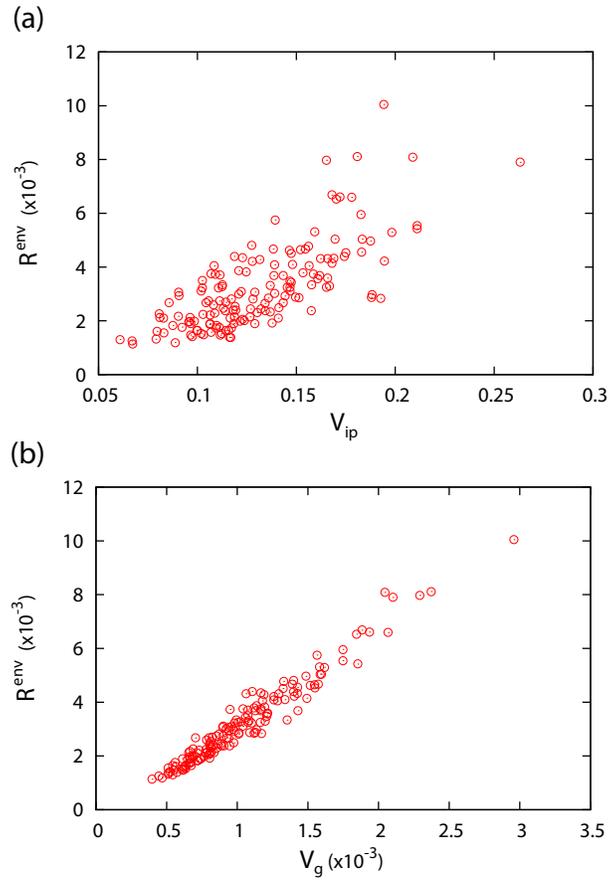}
\end{center}
\caption{
Supplementary Figure 1. The relationships of $R^{env}$ to (a) $V_{ip}$ and (b) $V_g$. The variances are computed by using the network and environment at generation 0 (before the environmental change). For the details of environmental changes to calculate $R^{env}$, see Simulation methods. 
}
\end{figure}

\begin{figure}[htbp]
\begin{center}
\includegraphics[width=10cm]{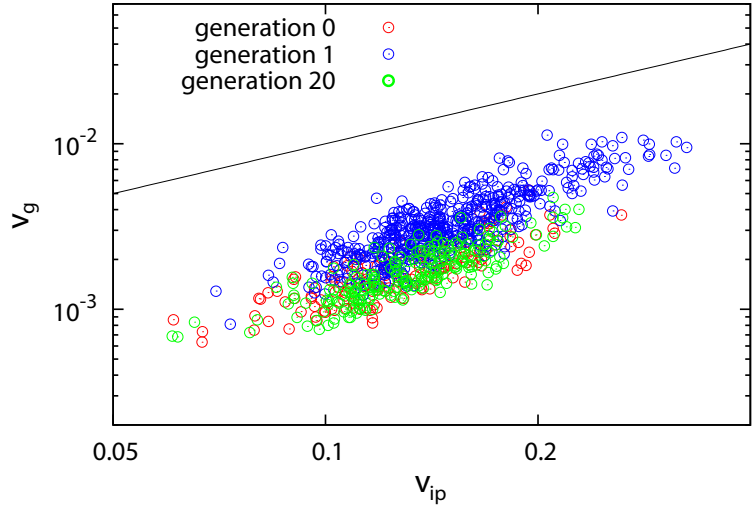}
\end{center}
\caption{
Supplementary Figure 2. The relationship between $V_{ip}$ and $V_g$ after the environmental change. 
$V_{ip}$ and $V_g$ before the environmental change (generation 0), immediately after the environmental change (generation 1), and after the adaptive evolution (generation 20) are plotted. After the environmental change, both $V_{ip}$ and $V_g$ increase , and then recover the original levels after 20 generations of the evolution. The solid line is $y=x$ for reference.
~\\
~\\
}
\end{figure}

\begin{figure}[htbp]
\begin{center}
\includegraphics[width=7cm]{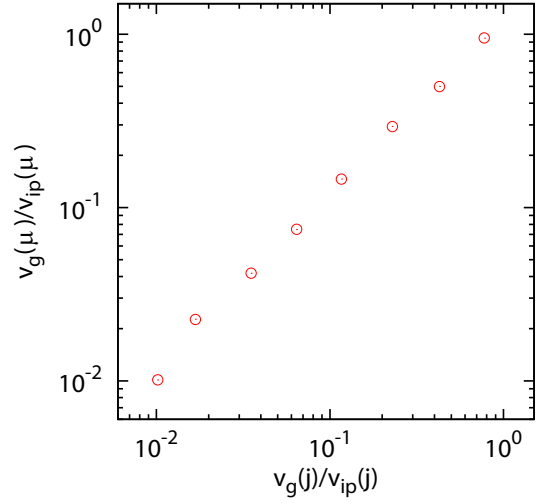}
\end{center}
\caption{
Supplementary Figure 3. The relationship between the ratio of $V_g(j)/V_{ip}(j)$ and $V_g(\mu)/V_ip(\mu)$. 
The variance ratio $V_g(j)/V_{ip}(j)$ is calculated by the least square method for all components. 
The data points are obtained with $m=10^{-6} \times 2^{\ell}$ for $\ell =1,2,\cdots,8$. 
}
\end{figure}

\end{document}